\documentclass[twocolumn, switch]{article}
\usepackage{preprint}
\usepackage{algorithm}
\usepackage{algorithmic}
\usepackage{amsmath, amsthm, amssymb, amsfonts}

\usepackage[numbers,square]{natbib}
\usepackage[utf8]{inputenc}
\usepackage[T1]{fontenc}
\usepackage{xcolor}
\usepackage[colorlinks = true,
            linkcolor = purple,
            urlcolor  = blue,
            citecolor = cyan,
            anchorcolor = black]{hyperref}
\usepackage{booktabs}
\usepackage{nicefrac}
\usepackage{microtype}
\usepackage{lineno}
\usepackage{float}
\usepackage{lipsum}
\usepackage{newfloat}
\DeclareFloatingEnvironment[name={Supplementary Figure}]{suppfigure}
\usepackage{sidecap}
\sidecaptionvpos{figure}{c}

\usepackage{titlesec}
\titlespacing\section{0pt}{12pt plus 3pt minus 3pt}{1pt plus 1pt minus 1pt}
\titlespacing\subsection{0pt}{10pt plus 3pt minus 3pt}{1pt plus 1pt minus 1pt}
\titlespacing\subsubsection{0pt}{8pt plus 3pt minus 3pt}{1pt plus 1pt minus 1pt}

\usepackage{eso-pic}
\usepackage{tikz}
\usepackage{xcolor}
\PassOptionsToPackage{colorlinks=true,linkcolor=gray,urlcolor=gray}{hyperref}

\usepackage{titling}
\usepackage{footmisc}
\usepackage{multirow}
\usepackage{amsmath}
\newtheorem{theorem}{Theorem}[section]
\newtheorem{proposition}[theorem]{Proposition}

\newcommand{\Author}[2]{\textbf{#1}\textsuperscript{#2}}

\title{PathRWKV: Enhancing Whole Slide Image Inference with Asymmetric Recurrent Modeling}

\author{
  \Author{Tianyi Zhang}{1} \and
  \Author{Sicheng Chen}{2,3}\and
  \Author{Borui Kang}{1}\and
  \Author{Dankai Liao}{3} \and
  \Author{Qiaochu Xue}{1} \and
  \Author{Bochong Zhang}{1}\and
  \Author{Fei Xia}{2} \and
  \Author{Zeyu Liu}{3} \and
  \Author{Yueming Jin}{1}
}

\date{
  \textsuperscript{1}Department of Electrical \& Computer Engineering, National University of Singapore \\
  \textsuperscript{2}Nhu Department of Electrical Engineering and Computer Science, University of California, Irvine \\
  \textsuperscript{3}PuzzleLogic Pte Ltd, Singapore 229594, Singapore \\
  [1em]
  \footnotesize \textbf{Corresponding author:} Fei Xia\texttt{<fei.xia@uci.edu>}, Zeyu Liu\texttt{<zeyuliu@puzzlelogic.com>}, Yueming Jin\texttt{<ymjin@nus.edu.sg>} \\
}

\begin{document}
\twocolumn[\begin{@twocolumnfalse}
\maketitle
\thispagestyle{empty}
\begin{abstract}
Whole Slide Imaging (WSI) has become a gold standard in cancer diagnosis, inspecting multi-scale information from cellular to tissue levels. Processing an entire WSI directly is infeasible due to GPU memory constraints; thus, Multiple Instance Learning (MIL) has emerged as the standard solution by partitioning WSIs into tiles. While recent two-stage MIL frameworks partially achieve memory efficiency by decoupling tile-level extraction from slide-level modeling, they still face four limitations:
(1) the conflict between training throughput and inference memory efficiency,
(2) the high susceptibility to overfitting on small-scale WSI datasets with sparse supervision,
(3) the disruption of spatial structural integrity during sampling-based training, and
(4) the inadequate modeling of multi-scale feature interactions within long sequences. We therefore introduce PathRWKV, a novel State Space Model designed for efficient and robust WSI analysis.
To resolve the computational trade-off, we propose an asymmetric structure utilizing max pooling aggregation, enabling parallelized training for high throughput and recurrent inference with constant ($\mathcal{O}(1)$) memory complexity.
To mitigate overfitting, we employ random sampling to enhance data diversity, with a multi-task learning module to regularize feature learning on limited data.
To restore spatial context, we introduce 2D sinusoidal position encoding to perceive the relative locations of tissue tiles.
To capture comprehensive representations, we integrate TimeMix and ChannelMix modules, enabling dynamic multi-scale feature modeling across temporal and spatial dimensions.
Experiments on 29,073 WSIs across 11 datasets demonstrate that PathRWKV outperforms 11 state-of-the-art methods on 10 datasets, establishing it as a scalable and solution with application potential.
\end{abstract}

\keywords{Whole slide image analysis \and multiple instance learning \and multi-task learning \and state space model}
\vspace{0.35cm}
\end{@twocolumnfalse}]
\section{Introduction}
Pathology diagnosis plays an essential role in clinical practice, leveraging the analysis of pathological images~\cite{cpia, wsi} to ensure accurate cancer diagnosis and treatment planning. The process begins with tissue biopsy/grossing by a specialist, and sample preparation workflows digitize these samples. This creates gigapixel-scale Whole Slide Images (WSIs) capturing both cell-level and tissue-level morphological details~\cite{wsi}. 
While WSIs contain rich, high-dimensional, multi-scale features, their colossal sizes make manual review labor-intensive and require highly specialized expertise, leading to inconsistent diagnoses across sites that could negatively impact the quality of healthcare~\cite{digital, cam16}. 
Deep learning--driven computational pathology techniques have emerged to ease pathologists’ burden and promote high-quality diagnosis by automatically identifying critical patterns in WSIs~\cite{clinical, clam, transpath}. Recent studies have further advanced this by quantifying pathologists' visual patterns to integrate expert cognitive strategies into diagnostic models, thereby aiming to minimize workload while maintaining precision~\cite{visual_patterns}. Nevertheless, the complex, multi-scale nature of WSIs sets challenges for deep learning models to capture and integrate features robustly across different scales~\cite{shuffle, deep}.

Specifically, end-to-end training on raw, high-resolution WSIs remains infeasible due to GPU memory constraints and extreme dimensionality~\cite{transpath, clinical}. Some studies downsample WSI to a thumbnail, while reduce computational complexity, this incurs significant information loss by discarding high-resolution details (e.g., cell morphology) critical for assessing disease progression~\cite{downsample, downsample2}. Consequently, most pipelines predominantly adopt Multiple Instance Learning (MIL) as a practical two-stage solution~\cite{abmil}. 
This approach breaks WSIs into smaller tiles (e.g., $224\times224$ pixels from an $80,000\times60,000$-pixel slide) and encodes them into dense feature representations using a foundation model (e.g., Prov-GigaPath~\cite{gigapath}). Subsequently, a slide-level backbone aggregates these compressed tile features to generate slide-level predictions~\cite{clam, transmil, dtfdmil}. This paradigm robustly enhances performance by enabling the processing of a larger number of tiles per iteration and leveraging the robust prior knowledge embedded in the foundation model. Accordingly, these advantages contribute to the high accuracy achieved by modern MIL frameworks~\cite{advances, visual_patterns}. However, critical challenges remain unresolved as detailed below:

A primary challenge in WSI analysis arises from the drastic variation in the number of tiles per slide due to diverse tissue dimensions~\cite{advances}. 
During training, a fixed number of tiles (e.g., 2,000) is uniformly sampled from each WSI to leverage a larger batch size. This maximizes GPU parallelization efficiency, and stabilizes gradient descent, thereby enhancing overall performance~\cite{transpath}. 
Conversely, during inference, to cover all regions for diagnostic evidences, the batch size is generally set to one to process all available tiles~\cite{clinical}. 
However, given the immense disparity in WSI dimensions (e.g., from 1,000 to over 40,000 tiles in the CAMELYON16 dataset~\cite{cam16} at 0.5 mpp with a 224$\times$224 patch size), processing large-scale slides can easily exceed GPU memory limits, particularly when deploying recent effective yet complex Transformers on resource-constrained edge devices~\cite{longnet}. This necessitates a slide-level structure that possesses parallel computing capabilities during training to handle large batch sizes, while retaining sequential processing efficiency during inference to model entire WSIs with minimal memory overhead~\cite{rwkv}.

Another critical challenge stems from the severe data inefficiency in WSI analysis, where high-capacity models struggle to generalize under the dual constraints of data scarcity and sparse supervision~\cite{dcpn_fewshot, pathology}. Specifically, obtaining annotated cohorts is prohibitive, often restricting datasets to limited sizes (e.g., fewer than 3,000 slides)~\cite{cpia}. This scarcity is exacerbated by the weak nature of slide-level labels, which provide supervision for only a fraction of the gigapixel-resolution tissue information~\cite{clinical}. Consequently, despite the improved feature representations from foundation models~\cite{gigapath, transpath}, the downstream aggregators remain prone to overfitting. Notably, complex structures like TransMIL~\cite{transmil} frequently underperform compared to simpler baselines (e.g., CLAM~\cite{clam}) in such low-data regimes~\cite{dcpn_fewshot}. This necessitates strategies that maximize the utility of limited slide-level data to enhance model generalization~\cite{cellmix, task}.

Furthermore, the disruption of spatial structural integrity during sampling-based MIL training significantly impedes the performance of permutation-variant methods. Conventional mini-batch training requires a fixed input size, and current methods primarily address this by selecting a subset of tiles to represent the whole WSI~\cite{abmil}. This inevitably disrupts the global spatial context. While this issue is negligible for permutation-invariant Attention-based methods (e.g., CLAM~\cite{clam}), it poses a critical challenge for the current permutation-variant state-of-the-art (SOTA) paradigms, like Transformers (e.g., TransMIL~\cite{transmil}) and State Space Models (SSMs) (e.g., MambaMIL~\cite{mambamil}). All of them rely on explicit modeling of spatial relationships and contextual dependencies. Consequently, the spatial information loss induced by sampling severely compromises the modeling capabilities of these structures~\cite{improving}. It is imperative to develop a mechanism that can effectively recover this missing spatial context~\cite{puzzletuning}.

The final challenge arises as most current methods face difficulties in adequately handling multi-scale feature interactions within long sequences. Effective diagnosis relies on the complementarity between fine-grained details (e.g., cell nuclei at 40$\times$) and coarse-grained context (e.g., tissue structure at 4$\times$), and the inherent ambiguity of cellular structures theoretically necessitates fuzzy logic or high-order topological modeling~\cite{tfuzz_fuzzy}. This requires a multi-scale modeling capability to simultaneously handle the relationships between fine-grained local homogeneity and coarse-grained global heterogeneity~\cite{cellmix}. However, existing methods demonstrate an inability to conduct effective multi-scale feature analysis for slide-level conclusions. Lightweight Attention-based methods benefit from simple, permutation-invariant structures but lack the capacity to capture intricate fine-grained relationships (e.g., inter-cell interactions). Conversely, while large Transformers excel at modeling local details, their generic structures often struggle to effectively align and fuse these heterogeneous multi-scale features into a unified slide-level representation. A promising solution is to model features from multiple perspectives, incorporating diverse indicators to construct a robust multi-scale understanding.

\begin{figure}[t]
\centerline{\includegraphics[width=\columnwidth]{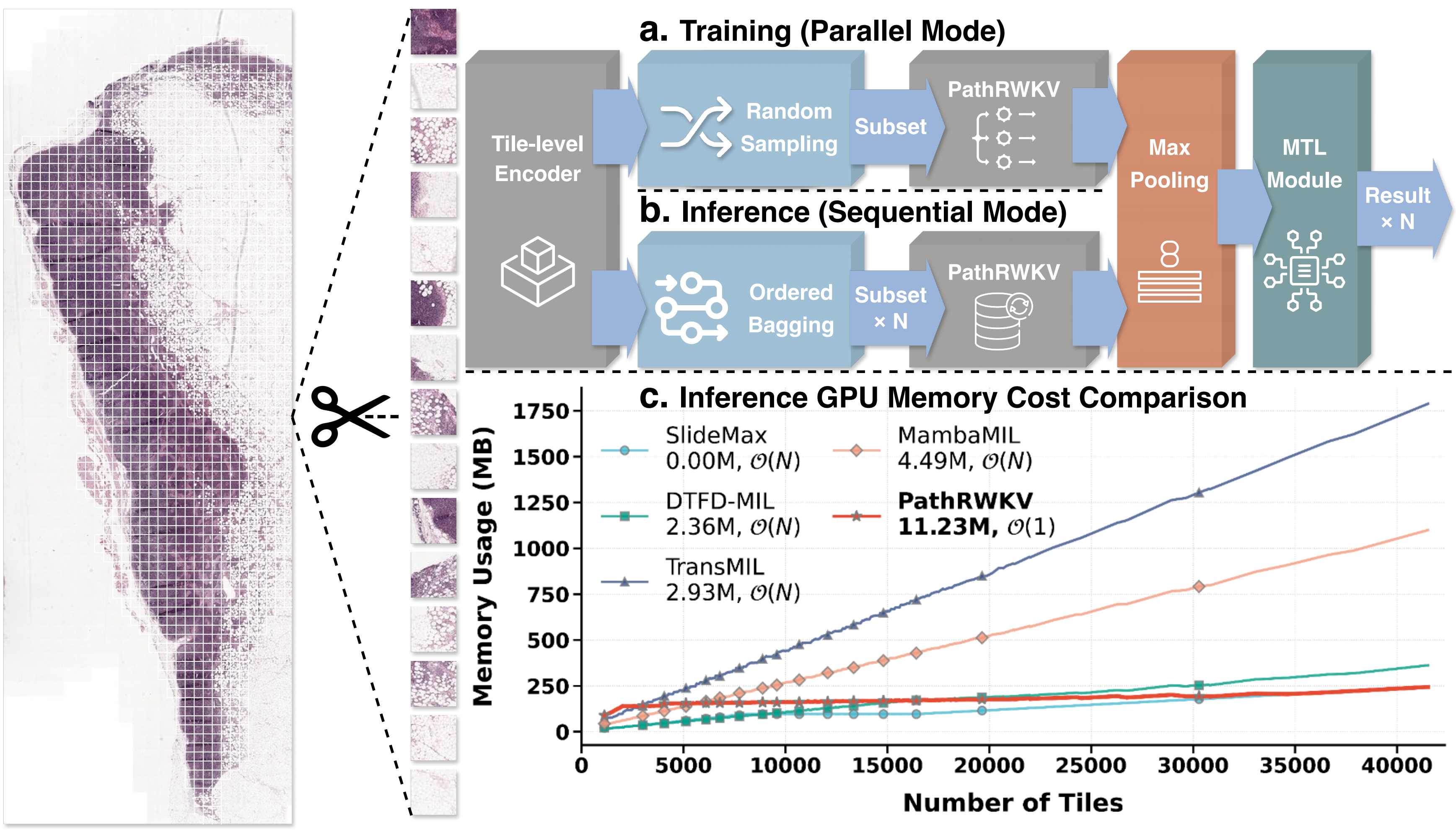}}
\caption{The asymmetric structure of PathRWKV and its effectiveness. (a) Parallel mode: The model samples a fixed maximum number of WSI tiles for multi-sample parallel processing to maintain efficiency during training. (b) Sequential mode: The model processes all WSI tiles for precise inference. It splits tiles into equal-sized bags and processed sequentially in a single forward pass. A memorable state retains and propagates information from prior bags. (c) GPU memory usage comparison during inference. PathRWKV maintains the $\mathcal{O}(1)$ spatial complexity, showing superior memory efficiency on long-context modeling compared with previous methods.}
\label{fig:asymmetric}
\end{figure}

Due to the similar requirements of long context modeling and multi-scale understanding between natural language processing (NLP) and MIL, previous MIL approaches based on NLP structures have proven effective (e.g., TransMIL~\cite{transmil} from Transformer, MambaMIL~\cite{mambamil} from Mamba). Among them, RWKV~\cite{rwkv} stands out by uniquely combining the efficient parallelizable training of Transformers with the linear complexity inference of RNNs. This makes it exceptionally suitable for processing the massive, sequential feature representations inherent in WSI analysis. Motivated by these properties, we propose PathRWKV, a time-decayed SSM tailored for efficient and robust WSI analysis.

To resolve the asymmetric memory and efficiency constraints, we propose an asymmetric structure (Fig.~\ref{fig:asymmetric}) that integrates max pooling aggregation with linear attention. This design enables a seamless transition between Transformer-like parallelization for high-throughput training and RNN-like sequential processing for inference, achieving constant ($\mathcal{O}(1)$) memory complexity regardless of slide size. Consequently, it achieves high throughput during training while ensuring exceptional memory efficiency during inference.

To mitigate overfitting and data scarcity inherent in weak supervision, we adopt a random sampling strategy with the multi-task learning (MTL) module. Random sampling acts as a dynamic data augmentation technique, counteracting the inductive bias of deterministic sampling and exposing the model to diverse subsets of tissue regions. The MTL module introduces auxiliary supervision signals to regularize feature space, preventing the model from memorizing noise in limited training samples. Together, these strategies exploit the potential of limited annotations and bolster model generalizability by capturing intrinsic inter-task dependencies.

To address the disruption of spatial structural integrity caused by random sampling, we leverage 2D sinusoidal position encoding (2D PE) to embed unique coordinate-based information into each tile feature. It is critical for the permutation-variant PathRWKV to recognize relative positions and reconstruct spatial relationships. This design effectively equips the model to preserve global spatial context, bridging the significant distributional gap between the stochastic bag-of-tiles input used during training and the ordered, sequential slide processing required for inference.

To tackle complex multi-scale feature interactions, we incorporate TimeMix and ChannelMix modules. The TimeMix module focuses on capturing long-range spatial dependencies and local homogeneity across the sequence of tiles, while the ChannelMix module focuses on high-level abstract semantic patterns. By jointly modeling these dimensions, the structure ensures a robust representation that encompasses both fine-grained cellular details and coarse-grained global tissue heterogeneity across the entire slide.

This work makes the following contributions:
\begin{itemize}
    \item We propose \textit{PathRWKV}, a novel SSM for efficient and robust slide-level modeling in computational pathology.
    
    \item We design an asymmetric slide-level structure that combines max pooling aggregation, enabling efficient parallelized training and recurrent inference with constant ($\mathcal{O}(1)$) memory complexity. Built upon this structure, we further introduce random sampling strategy and MTL module to mitigate overfitting under weak supervision and improve data efficiency and generalization.
    
    \item We restore spatial context by incorporating 2D PE, and enhance multi-scale representation learning via TimeMix and ChannelMix modules, enabling dynamic interaction between fine-grained cellular features and coarse-grained tissue structures.
    
    \item We conduct extensive experiments on 29,073 WSIs across 11 public datasets, demonstrating SOTA performance on 10 datasets. Beyond accuracy, these results highlight PathRWKV's potential as a scalable and reliable framework for large-scale pathological analysis, establishing a new perspective where the asymmetry between training and inference serves as a powerful inductive principle for computational pathology. The code is publicly available at \underline{https://github.com/Puzzle-Logic/PathRWKV}.
\end{itemize}

\section{Related Work}
The two-stage MIL paradigm becomes the mainstream recently, with the first stage extracts tile-level features, and the second aggregates them for slide-level predictions. Initial attempts within this paradigm employed simple aggregation strategies, such as average pooling (SlideAve) and max pooling (SlideMax) from MINNs~\cite{minns}, to combine tile features into a slide-level representation. While computationally efficient, these methods simply treat all tiles embeddings equally or focus exclusively on the most salient one, often failing to capture the complex, fine-grained information required for accurate diagnosis. To address this limitation, ABMIL~\cite{abmil} introduced a gated attention mechanism that adaptively weights tiles to enable instance-level interpretability. This simple yet effective mechanism has been widely adopted by most modern methods. Building on this, CLAM~\cite{clam} imposes instance clustering constraints to encourage diverse and discriminative feature learning, thereby improving model generalization. DSMIL~\cite{dsmil} further incorporates contrastive learning by combining instance- and bag-level supervision to better distinguish informative tiles. To address the challenge of limited data scale, DTFD-MIL~\cite{dtfdmil} proposes a double-tier feature distillation framework that utilizes pseudo-bags to virtually expand the training set, enhancing robustness in small-sample scenarios. Despite their widespread adoption, these attention-based methods often struggle to fully capture complex spatial dependencies within WSIs due to their inherent permutation invariance.

Transformers have been introduced to alleviate the permutation invariance of attention-based methods and improve global context modeling. A representative method, TransMIL~\cite{transmil}, employs a Transformer-based structure that explicitly encodes positional and structural relationships among tiles, enabling more effective global spatial reasoning. Furthermore, Prov-GigaPath~\cite{gigapath} enhances global information flow through a dilated attention mechanism, leveraging efficient sequence modeling structures like LongNet~\cite{longnet} to improve scalability. Despite these advancements, Transformers still face challenges regarding overfitting and high memory consumption, especially when trained on small-scale datasets.

State Space Models (SSMs) have emerged as a compelling alternative, balancing the efficiency of conventional attention methods with the long-range modeling capabilities of Transformers. S4MIL~\cite{s4mil} introduces the Structured State Space Sequence (S4) model to capture long-range dependencies across tiles. By imposing a structured state representation, it mitigates the overfitting risks associated with data-hungry Transformers. Building on this, MambaMIL~\cite{mambamil} integrates the selective state space model, Mamba, into the MIL pipeline, enabling linear scaling and selective information flow across tiles. Similarly, MamMIL~\cite{mammil} adapts the Mamba structure to model WSIs as long sequences, effectively capturing bidirectional contextual dependencies with minimal computational overhead. These SSM-based approaches not only offer superior scalability but also enhance generalization through their inherent inductive biases, making them particularly well-suited for WSI analysis on limited datasets. However, most current SSM-based approaches retain attention-based pooling mechanisms for final aggregation. This design necessitates storing features from all tiles, reintroducing an $\mathcal{O}(N)$ memory bottleneck during inference that undermines the inherent linear efficiency of the SSM backbone.
\section{Methods}
\label{sec:methods}
\begin{figure*}[!t]
\centerline{\includegraphics[width=\textwidth]{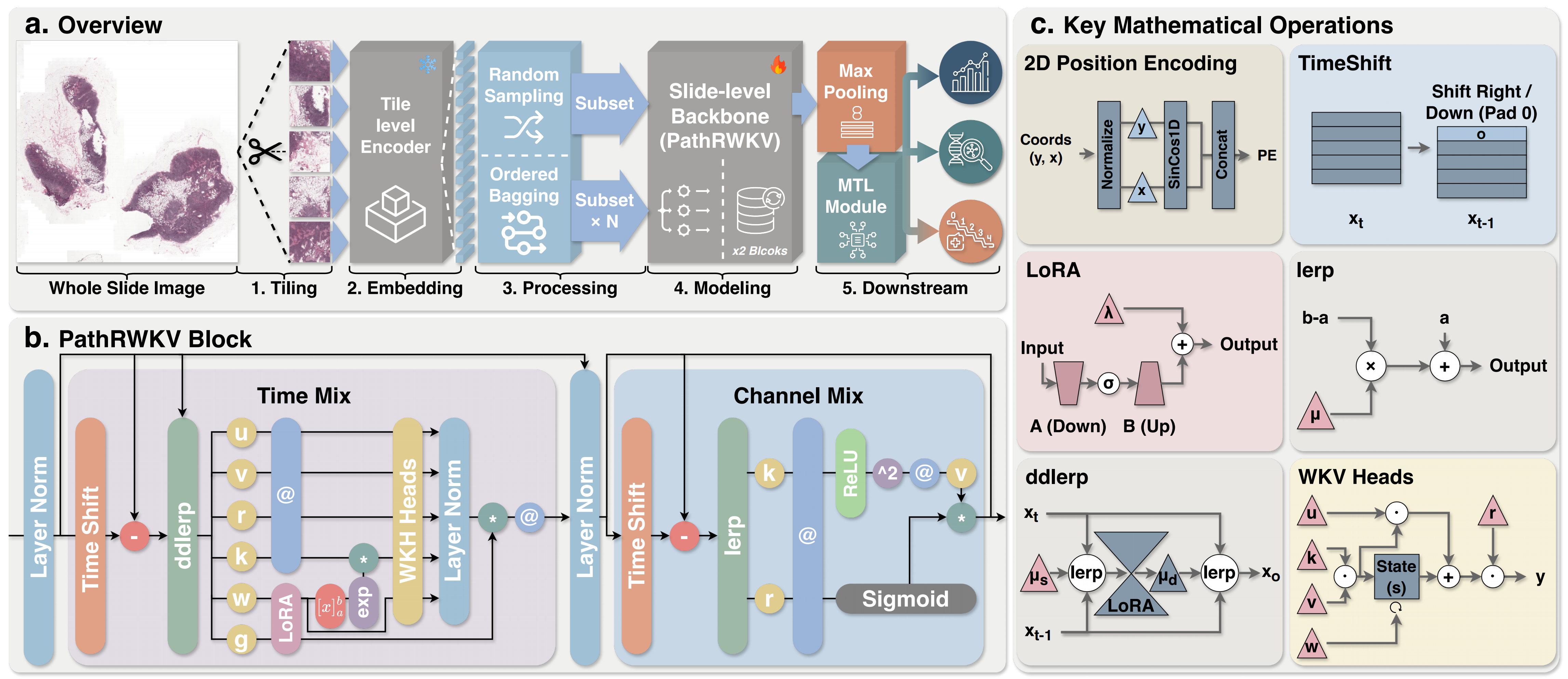}}
  \caption{Overview of PathRWKV. a) The pipeline begins with WSI tiling and tile-level feature embedding via Prov-GigaPath, followed by the slide-level backbone via PathRWKV, which enables multi-task learning for different downstream tasks. b) The PathRWKV block consists of the TimeMix module, which integrates tile features with previous states from multi-scale using linear attention, and the ChannelMix module, which blends features via spatial aspect. c) Details of the specific mechanisms employed, including 2D Position Encoding, TimeShift, LoRA, lerp, ddlerp, and WKV Heads.}
  \label{fig:main}
\end{figure*}

\subsection{Overview of the MIL Pipeline}
Fig.~\ref{fig:main}a illustrates the overall PathRWKV pipeline. Following existing works \cite{unpuzzle}, each WSI $S$ is first loaded at a target resolution (e.g., 0.5 microns per pixel (mpp)) and partitioned into a non-overlapping grid of tiles $\{T_{\text{i,j}}\}$ of size $T_{\text{size}}$. To ensure data quality, a two-stage filtering protocol is employed. First, tiles with tissue coverage below a predefined threshold (e.g., $<50\%$ of the tile area) are discarded. Second, tiles with pixel variance falling below a quantitative cutoff (e.g., $Var(I)<0.01$, where $I$ denotes the normalized pixel intensity in [0,1]) are removed. This ensures that only informative tiles are retained for downstream tasks.
After pre-processing and filtering, each tile is embedded into a dense semantic feature vector (Fig. \ref{fig:main}a) using a pathological foundation model (e.g., Prov-GigaPath \cite{gigapath}). 
This embedding process enhances both the training efficiency and performance of the slide-level MIL. 
During training, PathRWKV processes a randomly shuffled subset of tiles to facilitate efficient multi-slide learning; conversely, during inference, it utilizes the complete sequential tile sequence from the WSI. 
Finally, the slide-level output features from PathRWKV are passed to the MTL module to generate predictions for each task (e.g., cancer subtyping, tumor grading, overall survival).

\subsection{The PathRWKV Slide-level Backbone}
PathRWKV serves as the backbone for slide-level feature modeling, consisting of 2 blocks with a hidden dimension of 768, and 12 heads. The input tile features are first combined with 2D sinusoidal Position Encoding (2D PE) to restore spatial relationships disrupted during the sampling process. Subsequently, these features are processed by the PathRWKV blocks (Fig.~\ref{fig:main}b). 
Each block comprises a TimeMix module and a ChannelMix module, integrated with layer normalization and residual connections. 
The TimeMix module dynamically captures multi-scale inter-tile dependencies via the temporal dimension, while the ChannelMix module focuses on intra-tile feature interactions across the channel dimension. 
Output features from the final PathRWKV block are aggregated via max pooling to produce the final slide-level representation.

Fig.~\ref{fig:main}c illustrates the key mathematical operations within the TimeMix and ChannelMix modules. The TimeMix module is specifically designed to capture multi-scale temporal dependencies among tiles. 
It effectively integrates fine-grained local interactions via token shifting and interpolation with coarse-grained global context via time-decayed linear attention, combining the strengths of Transformers~\cite{vit} and RNNs~\cite{lstm}.

To capture short-range, fine-grained dependencies between adjacent tiles, the module first employs a token-shift mechanism, TimeShift. A shifted version of the input $x$, denoted as $x_{last}$, is generated using zero-padding:
\begin{equation}
    x_{last, t} = x_{t-1}, \quad x_{last, 0} = \mathbf{0}
\end{equation}
The current input $x$ is then mixed with $x_{last}$ via data-dependent linear interpolation (ddlerp). Unlike static interpolation, ddlerp dynamically computes the mixing coefficient $\mu$ using a Low-Rank Adaptation (LoRA) mechanism:
\begin{equation}
\begin{aligned}
    \delta x_t &= x_{last, t} - x_t \\
    x'_{t} &= x_t + \delta x_t \odot \mu_x \\
    \mu_t &= \lambda + \tanh(x'_t W_A) W_B \\
    x_{ddlerp, t} &= x_t + \delta x_t \odot \mu_t
\end{aligned}
\end{equation}
This mechanism allows the model to adaptively aggregate local information from the immediate predecessor based on the current context, ensuring that high-frequency local variations are preserved before global processing.

Following local aggregation, the interpolated features are projected into five vectors: receptance $r$, key $k$, value $v$, time-decay $w$, and gate $g$. To capture long-range, coarse-grained dependencies across the entire slide sequence, we employ a time-decayed linear attention mechanism. Notably, the decay rate $w$ is modulated by a LoRA projection, allowing for data-dependent decay speeds that can adaptively focus on relevant historical context. The global linear attention is computed efficiently as:
\begin{equation}
    y_t = r_t \odot \left( \sum_{i=1}^{t-1} \left( \prod_{j=i+1}^{t} w_j \right) \odot k_i v_i^\top + u \odot k_t v_t^\top \right)
    \label{eq:parallel}
\end{equation}
Crucially, this formulation can be switched to a recurrent structure, which underpins our asymmetric design. By maintaining a recurrent state $S_t$, the model propagates global context sequentially:
\begin{equation}
\begin{aligned}
    S_t &= w_t \odot S_{t-1} + k_t v_t^\top \\
    y_t &= r_t \odot (S_{t-1} + u \odot k_t v_t^\top)
\end{aligned}
\label{eq:sequential}
\end{equation}
Finally, the output is gated by $g$ and projected by $W_o$:
\begin{equation}
    y = W_o(\text{GroupNorm}(\mathbf{y}) \odot g)
\end{equation}
The ChannelMix module focuses on intra-tile feature interactions. Specifically, it employs learnable linear projections ($r, k, v$) to blend information across the channel dimension $D$ for each tile independently. This is coupled with a Squared ReLU activation:

\begin{equation}
    \sigma(x) = \max(0, x)^2
\end{equation}

which induce robust non-linear transformations, enabling the extraction of complex morphological features within each tile.

\subsection{Asymmetric Structure and Max Pooling Aggregation}
As illustrated in Fig.~\ref{fig:asymmetric}, distinct from the standard token-by-token processing in original RWKV, we propose a novel hybrid set-by-set recurrent architecture tailored for high-resolution WSI analysis. While leveraging the mathematical efficiency of linear attention, our core innovation lies in the asymmetric formulation of slide-level modeling to resolve the memory bottlenecks inherent in existing SSMs.

During the training phase, we adopt a set-based parallel strategy to maximize throughput. Instead of processing tiles sequentially, the model ingests a fixed number of sampled tiles as a dense batch. By utilizing a parallel CUDA kernel, we compute cumulative states and gradients simultaneously. This design fully exploits the massive parallelism of modern GPUs, facilitating rapid backpropagation and stable convergence compared to pure recurrent training.

During the inference phase, we introduce a streaming recurrent mechanism to achieve constant $\mathcal{O}(1)$ spatial complexity. A critical limitation in previous SSM-MIL methods (e.g., S4MIL, MambaMIL) is their reliance on Attention-based aggregation, which necessitates storing all tile features ($\mathcal{O}(N)$) for the final calculation, causing memory overflows on gigapixel slides. To overcome this, our architecture decomposes the WSI into sequential chunks. A latent state S (of size HeadSize$\times$HeadSize) acts as a memory carrier, propagating context across chunks via a recurrent kernel.

Furthermore, a critical bottleneck remains in the aggregation mechanism of existing frameworks, including recent SSMs. While SSM backbones theoretically allow constant spatial complexity ($\mathcal{O}(1)$) during inference, they typically adapt the Gated Attention mechanism from ABMIL~\cite{abmil} for slide-level aggregation. This approach necessitates computing a global Softmax normalization term across all tiles:

\begin{equation}
    \alpha_k = \frac{\exp(w^\top h_k)}{\sum_{i=1}^{N} \exp(w^\top h_i)}
\end{equation}

Consequently, the feature vectors of all $N$ tiles must be retained in GPU memory to calculate the denominator until the entire slide is processed, forcing the inference memory complexity to scale linearly with the slide size ($\mathcal{O}(N)$). This disrupts the memory efficiency gained by the SSM backbone.

To achieve a fully memory-efficient pipeline, we integrate the backbone's sequential processing with a streaming aggregation strategy. We decompose the WSI $\mathcal{X}$ into bags $\mathcal{B}_1, \mathcal{B}_2, \ldots, \mathcal{B}_M$ aligned with the inference chunks, and identify the \textit{feature-wise max operation} as the optimal choice due to its recursive update property. Given the tile encoder $\phi_\theta$, we define the local summary $z_i$ and the combination rule as:
\begin{equation}
z_i = g\bigl(\mathcal{B}_i\bigr):=\max_{x \in \mathcal{B}_i} \phi_\theta(x), \quad \mathrm{Comb}(a,b) = \max(a,b)
\end{equation}

We then update the slide-level representation sequentially:
\begin{equation}
h_i = \mathrm{Comb}(h_{i-1}, z_i), \quad h_0 = \varnothing
\label{eq:update}
\end{equation}
where $h_i$ represents the summary of the first $i$ bags. Unlike attention mechanisms, this design decouples memory usage from sequence length, achieving true $\mathcal{O}(1)$ space complexity during inference. This enables PathRWKV to process arbitrarily large slides on edge devices while preserving consistency with the ideal global computation. Comprehensive theoretical analysis, including proofs for the unbiased nature of the gradients and memory complexity comparisons, are provided in the supplementary materials.

\subsection{Random Sampling and 2D Position Encoding}
Given the high dimensionality of WSIs and the sparse supervision signal, models are prone to overfitting. To mitigate this, we employ the random sampling strategy during training. Instead of processing the entire slide or a fixed region, we randomly sample a subset of tiles from the WSI in each iteration. This approach acts as a strong data augmentation technique, preventing the model from memorizing specific tile sequences and enhancing its generalization capabilities.

However, a critical side effect of random sampling is the disruption of the intrinsic 2D spatial structure and the anatomical adjacency of the tissue microenvironment. This is the main advantage of recent permutation-variant methods (e.g., MmabaMIL~\cite{mambamil}) compared to early permutation-invariant methods (e.g., ABMIL~\cite{abmil}). To compensate for this loss of spatial context and enable PathRWKV to model geometry-aware dependencies, we introduce a 2D PE. Formally, let $\mathbf{z}_i \in \mathbb{R}^D$ denote the feature embedding of the $i$-th tile, and $(x_i, y_i)$ represent its normalized spatial coordinates. We employ sinusoidal functions to map these coordinates into a continuous embedding space:

\begin{equation}
\begin{aligned}
    \text{PE}(p, 2k) &= \sin\left(p / \Omega^{4k/D}\right) \\
    \text{PE}(p, 2k+1) &= \cos\left(p / \Omega^{4k/D}\right)
\end{aligned}
\end{equation}

where $\Omega$ is a scaling factor and $p \in \{x_i, y_i\}$. The final spatial embedding $\mathbf{P}_i$ is constructed by concatenating the encodings of horizontal and vertical coordinates and injected into the tile features via addition: $\hat{\mathbf{z}}_i = \mathbf{z}_i + \mathbf{P}_i$. This ensures that geometric relationships are restored regardless of the sampling order.

\subsection{Multi-task Learning}
Finally, to maximize the utility of limited annotated data and enhance training efficiency, we incorporate a multi-task learning (MTL) module. It comprises multiple prediction heads, allowing the model to learn from diverse clinical objectives simultaneously. By leveraging task-wise correlations, the model extracts more discriminative features, further reducing the risk of overfitting on the feature distribution of a single task. We employ Cross-Entropy, Cox proportional hazards, and L1 losses for classification, survival analysis, and regression tasks, respectively. The total loss is aggregated as $\mathcal{L}_{total} = \sum_{i=1}^{T} \mathcal{L}_i$, where $T$ represents the number of task heads. Crucially, gradients are computed only for tasks with available labels, enabling flexible training on partially annotated datasets.
\begin{table*} [htbp]
\centering
\caption{The performance comparison with SOTA methods on eleven downstream datasets. Statistical significance compared to the proposed PathRWKV is denoted using Welch's t-test: $^*$: $0.01<p<0.05$, $^{\dagger}$: $0.001<p<0.01$, $^{\ddagger}$: $p<0.001$.}
\label{tab:main}
\setlength{\tabcolsep}{1pt}
\resizebox{\textwidth}{!}{
\begin{tabular}{l|l|cccccccccccc}
\hline
\multirow{2}{*}{\textbf{Dataset}} & \multirow{2}{*}{\textbf{Metric}} & \textbf{SlideAve} & \textbf{SlideMax} & \textbf{ABMIL} & \textbf{CLAM} & \textbf{DSMIL} & \textbf{DTFD-MIL} & \textbf{TransMIL} & \textbf{GigaPath} & \textbf{S4MIL} & \textbf{MamMIL} & \textbf{MambaMIL} & \textbf{PathRWKV} \\
&& \textbf{\cite{minns}} & \textbf{\cite{minns}} & \textbf{\cite{abmil}} & \textbf{\cite{clam}} & \textbf{\cite{dsmil}} & \textbf{\cite{dtfdmil}} & \textbf{\cite{transmil}} & \textbf{\cite{gigapath}} & \textbf{\cite{s4mil}} & \textbf{\cite{mammil}} & \textbf{\cite{mambamil}} & \textbf{Ours} \\
\hline
\multirow{3}{*}{\textbf{PANDA}} & \textbf{Acc.[\%]} & 63.92$^{\ddagger}$ & 62.74$^{\ddagger}$ & 76.06$^{\dagger}$ & 75.97$^*$ & 76.39$^*$ & 75.53$^{\dagger}$ & 75.79$^*$ & 73.58$^{\dagger}$ & 75.93$^{\dagger}$ & 75.70$^{\dagger}$ & 75.99$^{\dagger}$ & \textbf{76.45} \\
 & \textbf{AUC[\%]} & 89.53$^{\ddagger}$ & 88.54$^{\ddagger}$ & 94.71$^{\dagger}$ & 94.74$^*$ & 94.20$^*$ & 93.93$^*$ & 93.87$^*$ & 93.67$^*$ & 94.07$^*$ & 94.81$^*$ & 94.30$^{\dagger}$ & \textbf{94.89} \\
 & \textbf{F1[\%]} & 60.76$^{\ddagger}$ & 55.94$^{\ddagger}$ & 70.30$^{\dagger}$ & 69.90$^*$ & 70.12$^*$ & 70.03$^*$ & 69.02$^{\dagger}$ & 68.06$^{\dagger}$ & 70.70$^{\dagger}$ & 69.94$^*$ & 70.56$^{\dagger}$ & \textbf{70.81} \\
\hline
\multirow{3}{*}{\textbf{CAMELYON16}} & \textbf{Acc.[\%]} & 68.99$^{\ddagger}$ & 72.87$^{\ddagger}$ & \textbf{98.45}$^*$ & \textbf{98.45}$^*$ & 92.25$^*$ & \textbf{98.45}$^*$ & \textbf{98.45}$^*$ & \textbf{98.45}$^*$ & \textbf{98.45}$^*$ & 97.67$^*$ & 97.67$^*$ & \textbf{98.45} \\
 & \textbf{AUC[\%]} & 55.19$^{\ddagger}$ & 74.44$^{\ddagger}$ & 97.90$^*$ & 98.77$^*$ & 96.51$^{\dagger}$ & 98.65$^*$ & 98.80$^*$ & 97.93$^{\dagger}$ & 98.92$^*$ & 98.67$^*$ & 99.03$^{\dagger}$ & \textbf{99.11} \\
 & \textbf{F1[\%]} & 62.70$^{\ddagger}$ & 70.29$^{\dagger}$ & \textbf{98.34}$^*$ & \textbf{98.34}$^*$ & 91.47$^{\dagger}$ & \textbf{98.34}$^*$ & \textbf{98.34}$^*$ & \textbf{98.34}$^*$ & \textbf{98.34}$^*$ & 97.52$^*$ & 97.50$^*$ & \textbf{98.34} \\
\hline
\multirow{3}{*}{\textbf{IMP-CRS-2024}} & \textbf{Acc.[\%]} & 92.33$^{\ddagger}$ & 91.67$^{\dagger}$ & 94.67$^*$ & 94.22$^{\dagger}$ & 94.44$^*$ & 94.33$^*$ & 94.44$^*$ & 94.56$^*$ & \textbf{94.78}$^*$ & 94.22$^{\dagger}$ & 94.33$^*$ & \textbf{94.78} \\
 & \textbf{AUC[\%]} & 98.59$^{\ddagger}$ & 98.56$^{\dagger}$ & 99.36$^{\dagger}$ & 99.42$^*$ & 99.40$^{\dagger}$ & 99.41$^{\dagger}$ & 99.44$^{\dagger}$ & 99.43$^{\dagger}$ & 99.44$^*$ & 99.41$^{\dagger}$ & 99.43$^*$ & \textbf{99.45} \\
 & \textbf{F1[\%]} & 92.67$^{\ddagger}$ & 92.08$^{\dagger}$ & 94.90$^*$ & 94.40$^{\ddagger}$ & 94.69$^*$ & 94.67$^*$ & 94.54$^{\dagger}$ & 94.83$^*$ & 94.85$^*$ & 94.41$^*$ & 94.52$^*$ & \textbf{95.09} \\
\hline
\multirow{3}{*}{\textbf{TCGA-BRCA}} & \textbf{Acc.[\%]} & 59.20$^*$ & 55.19$^*$ & 59.30$^{\dagger}$ & 59.39$^{\dagger}$ & 59.39$^{\dagger}$ & 59.30$^{\dagger}$ & 59.84$^{\dagger}$ & 59.30$^{\dagger}$ & 59.39$^{\dagger}$ & 59.30$^{\dagger}$ & 59.30$^*$ & \textbf{60.11} \\
 & \textbf{AUC[\%]} & 61.14$^{\dagger}$ & 56.97$^*$ & 63.42$^{\dagger}$ & 63.10$^*$ & 64.32$^*$ & 64.18$^*$ & 61.47$^{\dagger}$ & 64.12$^*$ & 63.71$^*$ & 59.84$^*$ & 59.67$^{\dagger}$ & \textbf{64.35} \\
 & \textbf{F1[\%]} & 25.82$^*$ & 26.09$^*$ & 27.54$^{\dagger}$ & 30.00$^*$ & 30.04$^*$ & 28.64$^*$ & 26.57$^*$ & 29.41$^*$ & 29.59$^{\dagger}$ & 28.73$^*$ & 25.88$^*$ & \textbf{30.40} \\
\hline
\multirow{3}{*}{\textbf{TCGA-GBM}} & \textbf{Acc.[\%]} & 98.60$^*$ & 97.90$^{\dagger}$ & 98.60$^*$ & 98.60$^*$ & 99.30$^{\dagger}$ & 98.60$^*$ & 97.90$^{\dagger}$ & 99.30$^*$ & 99.30$^*$ & 98.60$^*$ & 99.30$^{\dagger}$ & \textbf{100.00} \\
 & \textbf{AUC[\%]} & 66.27$^{\dagger}$ & 65.71$^*$ & \textbf{66.67}$^*$ & \textbf{66.67}$^{\dagger}$ & 66.51$^*$ & 66.19$^*$ & 65.87$^*$ & 66.59$^{\dagger}$ & \textbf{66.67}$^*$ & \textbf{66.67}$^*$ & \textbf{66.67}$^*$ & \textbf{66.67} \\
 & \textbf{F1[\%]} & 74.65$^*$ & 49.47$^*$ & 74.65$^{\dagger}$ & 74.65$^*$ & 74.65$^{\dagger}$ & 74.65$^*$ & 49.47$^*$ & 89.82$^{\dagger}$ & 89.82$^*$ & 74.65$^*$ & 89.82$^*$ & \textbf{100.00} \\
\hline
\multirow{3}{*}{\textbf{TCGA-LGG}} & \textbf{Acc.[\%]} & 68.81$^{\dagger}$ & 68.81$^{\dagger}$ & 68.14$^{\dagger}$ & 61.07$^*$ & 68.47$^*$ & 68.15$^{\dagger}$ & 68.44$^{\dagger}$ & 68.46$^{\dagger}$ & 66.44$^*$ & 68.50$^{\dagger}$ & 68.14$^*$ & \textbf{69.13} \\
 & \textbf{AUC[\%]} & 71.80$^{\dagger}$ & 71.67$^{\dagger}$ & 71.89$^*$ & 67.59$^*$ & 71.95$^{\dagger}$ & 71.79$^{\dagger}$ & 67.88$^{\dagger}$ & 71.68$^*$ & 72.10$^*$ & 71.68$^*$ & 71.89$^*$ & \textbf{72.61} \\
 & \textbf{F1[\%]} & 68.54$^{\dagger}$ & 68.00$^{\dagger}$ & 68.23$^{\dagger}$ & 60.28$^*$ & 68.46$^*$ & 68.38$^*$ & 66.44$^*$ & 62.35 & 63.89$^*$ & 68.18$^*$ & 66.69$^{\dagger}$ & \textbf{68.96} \\
\hline
\multirow{3}{*}{\textbf{TCGA-CESC}} & \textbf{Acc.[\%]} & 62.96$^*$ & 62.96$^{\dagger}$ & 59.26$^{\dagger}$ & 66.67$^*$ & \textbf{70.37}$^{\dagger}$ & 59.26$^*$ & 66.67$^*$ & 67.78$^{\dagger}$ & 55.56$^{\dagger}$ & 66.67$^{\dagger}$ & 59.26$^*$ & \textbf{70.37} \\
 & \textbf{AUC[\%]} & 68.13$^{\dagger}$ & 56.59$^{\dagger}$ & 62.09$^{\dagger}$ & 65.93$^{\dagger}$ & 70.88$^*$ & 68.13$^{\dagger}$ & 70.33$^*$ & 71.43$^*$ & 56.04$^{\ddagger}$ & 67.03$^*$ & 58.24$^{\ddagger}$ & \textbf{75.27} \\
 & \textbf{F1[\%]} & 62.50$^*$ & 62.91$^*$ & 59.03$^{\dagger}$ & 66.67$^*$ & 69.32$^{\dagger}$ & 58.35$^*$ & 66.48$^*$ & 67.75$^{\dagger}$ & 55.49$^{\dagger}$ & 65.92$^*$ & 58.35$^{\dagger}$ & \textbf{70.33} \\
\hline
\multirow{3}{*}{\textbf{TCGA-ESCA}} & \textbf{Acc.[\%]} & 47.22$^{\dagger}$ & 38.89$^{\dagger}$ & 47.22$^{\dagger}$ & \textbf{55.56}$^{\dagger}$ & 47.22$^{\dagger}$ & 47.22$^{\dagger}$ & 47.22$^*$ & \textbf{55.56}$^*$ & 50.00$^{\dagger}$ & 50.00$^*$ & \textbf{55.56}$^*$ & \textbf{55.56} \\
 & \textbf{AUC[\%]} & 65.00$^*$ & 48.57$^{\dagger}$ & 63.35$^*$ & 65.09$^{\dagger}$ & 64.39$^{\dagger}$ & 65.33$^*$ & 64.68$^{\dagger}$ & 65.35$^{\dagger}$ & 65.10$^{\dagger}$ & 65.09$^*$ & 65.47$^{\dagger}$ & \textbf{65.96} \\
 & \textbf{F1[\%]} & 38.67$^*$ & 28.94$^{\dagger}$ & 39.09$^{\dagger}$ & 37.20$^{\dagger}$ & 26.67$^*$ & 32.46$^{\dagger}$ & 41.44$^{\dagger}$ & 44.14$^{\dagger}$ & 43.52$^{\dagger}$ & 35.08$^{\dagger}$ & 32.22$^{\dagger}$ & \textbf{48.31} \\
\hline
\multirow{1}{*}{\textbf{TCGA-LIHC}} & \textbf{C-Index} & 0.522$^{\dagger}$ & 0.533$^*$ & 0.549$^{\dagger}$ & 0.546$^*$ & 0.553$^*$ & 0.581$^{\dagger}$ & 0.575$^*$ & 0.573$^*$ & 0.550$^{\dagger}$ & 0.529$^{\dagger}$ & 0.574$^*$ & \textbf{0.584} \\
\hline
\multirow{1}{*}{\textbf{TCGA-BLCA}} & \textbf{C-Index} & 0.472$^*$ & 0.561$^*$ & 0.573$^*$ & 0.572$^{\dagger}$ & 0.571$^{\dagger}$ & 0.567$^{\dagger}$ & 0.559$^{\dagger}$ & 0.572$^*$ & 0.576$^{\dagger}$ & 0.538$^{\dagger}$ & 0.575$^*$ & \textbf{0.579} \\
\hline
\multirow{1}{*}{\textbf{TCGA-LUNG}} & \textbf{C-Index} & 0.490$^{\dagger}$ & 0.534$^*$ & 0.537$^{\dagger}$ & 0.550$^{\dagger}$ & 0.518$^*$ & 0.535$^*$ & 0.543$^*$ & 0.505$^{\dagger}$ & 0.525$^{\dagger}$ & 0.529$^{\dagger}$ & \textbf{0.547}$^*$ & 0.518 \\
\hline
\end{tabular}
}
\end{table*}

\section{Experiments}
\subsection{Datasets and Downstream Tasks}
\begin{figure} [htbp]
\centerline{\includegraphics [width=\columnwidth]{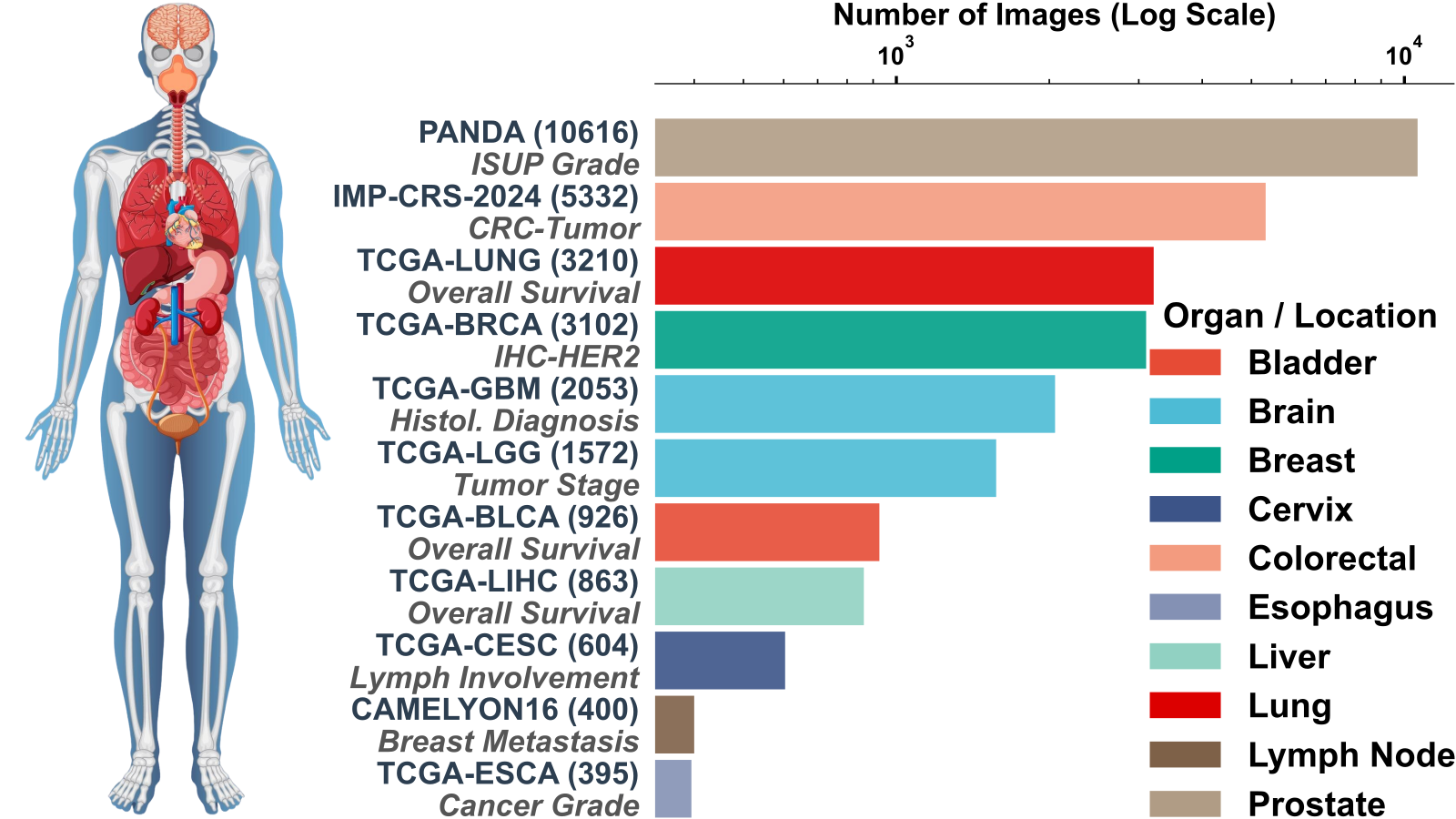}}
\caption{Summary of implemented datasets and clinical tasks. The horizontal bars display the number of WSIs for each cohort on a logarithmic scale, alongside their corresponding prediction targets. This collection covers a wide spectrum of sample sizes and clinical objectives, ensuring a comprehensive evaluation of model generalizability.}
\label{fig:dataset}
\end{figure}

We evaluated PathRWKV on 11 datasets across 9 downstream tasks covering diverse diagnostic scenarios, as shown in Fig.~\ref{fig:dataset}, to demonstrate its performance and generalizability.
The PANDA~\cite{panda} dataset with the ISUP Grade task assesses prostate cancer aggressiveness;
CAMELYON16~\cite{cam16} with the Breast Metastasis task classifies lymph nodes as normal or tumorous;
IMP-CRS-2024~\cite{crs2024} with the CRC-Tumor task identifies tumor tissues in colorectal images.
The TCGA~\cite{tcga} datasets cover multiple cancer types and tasks:
TCGA-BRCA with the IHC-HER2 task predicts HER2 receptor status from H\&E-stained slides;
TCGA-GBM with the Histological Diagnosis task classifies glioblastoma subtypes based on morphology;
TCGA-LGG with the Tumor Stage task predicts the WHO grade of lower-grade gliomas;
TCGA-CESC with the Lymphovascular Involvement task detects the presence of lymphovascular invasion;
TCGA-ESCA with the Cancer Grade task assesses the histological differentiation grade of esophageal carcinoma;
and TCGA-LIHC, TCGA-BLCA, and TCGA-LUNG, the combination of TCGA-LUAD and TCGA-LUSC, with the Overall Survival task predict patient survival time in months from liver, bladder, and lung tissue morphology, respectively.

\subsection{Implementation Details}
We use the UnPuzzle framework~\cite{unpuzzle} for preprocessing, where each WSI is tessellated into $224 \times 224$ patches at 0.5 mpp and embedded using the tile-level encoder of Prov-GigaPath~\cite{gigapath}. All models are initialized from scratch and trained for 100 epochs using the AdamW optimizer and a cosine decay scheduler with a final learning rate factor of 0.1. We conduct a grid search over ten learning rates ($1\times10^{-6}$ to $1\times10^{-3}$) and employ early stopping with a patience of 10 epochs based on validation loss. During training, we use a batch size of 4 and randomly sample a maximum of 2,000 tiles per WSI. For evaluation, we select the checkpoint with the lowest validation loss, use a batch size of 1, and process all tiles per WSI. We report the average performance of the top-3 results for each metric. For statistical reporting, significance was evaluated using Welch's t-test~\cite{ttest} across these top-3 performing learning rate configurations to compare each method against PathRWKV. All experiments were conducted on 4 NVIDIA RTX4090 GPUs using Python 3.12.12, PyTorch 2.9.1, and CUDA 12.8.

\subsection{Comparison with SOTA Methods}
To demonstrate the effectiveness of PathRWKV for slide-level WSI modeling, we compared it against 11 state-of-the-art (SOTA) methods, including SlideAve and SlideMax from MINNs~\cite{minns}, ABMIL~\cite{abmil}, CLAM~\cite{clam}, DSMIL~\cite{dsmil}, DTFD-MIL~\cite{dtfdmil}, TransMIL~\cite{transmil}, Prov-GigaPath~\cite{gigapath}, S4MIL~\cite{s4mil}, MambaMIL~\cite{mambamil}, and MamMIL~\cite{mammil}.

As presented in Tab.~\ref{tab:main}, PathRWKV demonstrates superior performance and robust generalizability, achieving SOTA results on 10 out of 11 datasets across 9 distinct downstream tasks. Specifically, on standard classification benchmarks such as CAMELYON16 and IMP-CRS-2024, most deep learning-based methods achieve high metrics ($>90\%$ Accuracy, AUC, and F1), with the exception of simple pooling strategies, SlideAve and SlideMax, that lack the representational capacity for comprehensive slide-level modeling. Among high-performing models, PathRWKV consistently secures the highest scores across all three metrics.

In contrast, the TCGA datasets present a significantly more challenging scenario, where the average performance of most methods drops to approximately 70\% due to the intrinsic complexity and heterogeneity of the samples. Despite these challenges, PathRWKV establishes its efficacy on the majority of TCGA datasets, validating the capability of the proposed TimeMix and ChannelMix modules to capture complex pathological dependencies. However, we observe a performance gap on the TCGA-LUNG overall survival task, where MambaMIL achieves the leading performance. We attribute this to the inherent trade-off of our max pooling strategy. While it ensures $\mathcal{O}(1)$ inference memory, it focuses on the most salient features and may inadvertently discard global contextual cues (e.g., total tumor burden) that are beneficial for specific prognostic predictions, which Attention mechanism in MambaMIL preserve better. Nevertheless, the strong performance of both PathRWKV and MambaMIL underscores the structural advantage of SSMs over conventional pooling models and Transformers in modeling multi-scale relationships within WSIs.

\begin{figure*} [t]
  \centerline{\includegraphics [width=\textwidth]{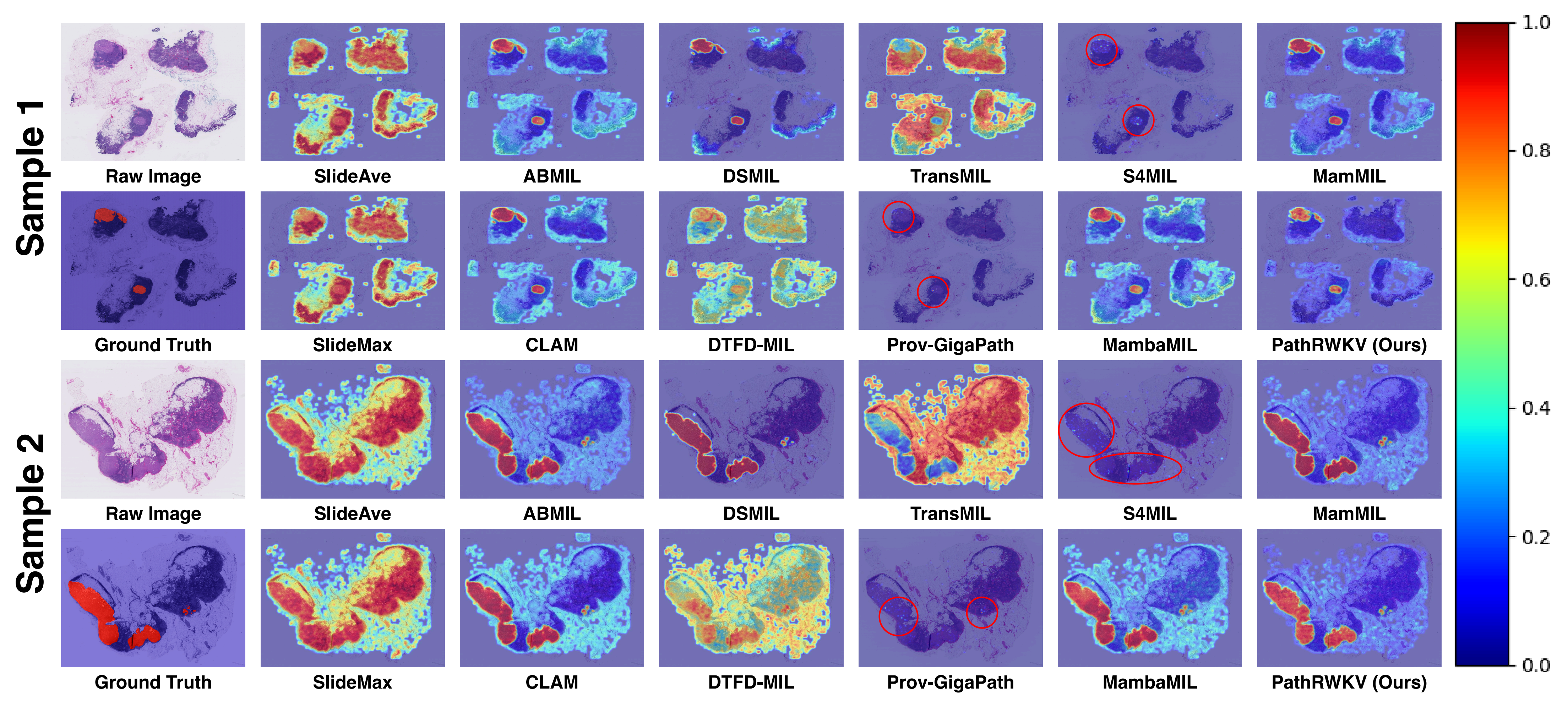}}
  \caption{Visualization of a high-grade lesion WSI sample from the CAMELYON16 dataset. From left to right: the raw image, the ground truth label annotated by the pathologist, feature embeddings extracted by the Prov-GigaPath tile-level encoder, and CAMs from each model.}
  \label{fig:cam}
\end{figure*}

\begin{figure} [t]
  \centerline{\includegraphics [width=\columnwidth]{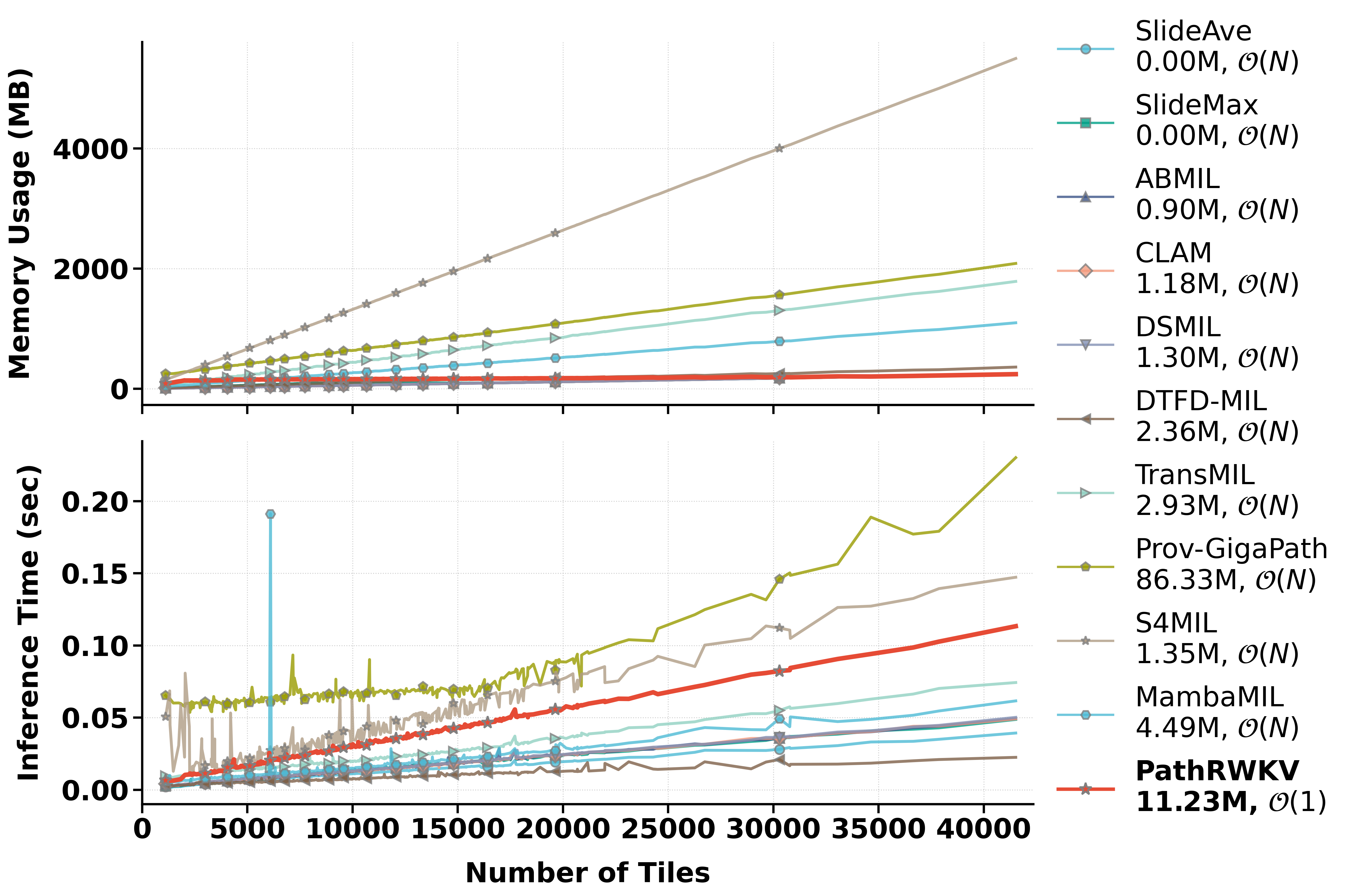}}
  \caption{GPU memory consumption and inference time versus the number of input tiles, where "M" denotes the number of parameters in millions. PathRWKV demonstrates superior efficiency with constant $\mathcal{O}(1)$ memory usage, significantly outperforming all baseline methods that exhibit linear $\mathcal{O}(N)$ memory growth, including SSMs with attention aggregation. Regarding inference speed, despite the inherent linear $\mathcal{O}(N)$ time complexity, PathRWKV achieves competitive and stable performance relative to its parameter size.}
  \label{fig:memory}
\end{figure}

A core innovation of PathRWKV is its asymmetric design, which fundamentally optimizes GPU memory utilization during inference. Fig.~\ref{fig:memory} compares the memory consumption profiles of various methods as the number of input tiles increases. Conventional attention-based methods (e.g., ABMIL, CLAM), Transformers (e.g., TransMIL, Prov-GigaPath), and even recent linear-complexity models (e.g., MambaMIL) exhibit linear memory growth ($\mathcal{O}(N)$). Although these SSM-based approaches utilize linear attention mechanisms similar to PathRWKV, they typically rely on the Gated Attention mechanism from ABMIL~\cite{abmil} for final aggregation. This strategy requires matrix multiplication between the complete input and output tensors to compute attention scores, necessitating the retention of the entire input tensor in GPU memory until the output is generated. This dependency causes memory usage to scale linearly with sequence length, effectively negating the inherent efficiency advantages of the SSM backbone and severely constraining applicability to large-scale WSIs in resource-limited environments. In contrast, by equal contribution of linear attention and max aggregation strategy, PathRWKV achieves constant memory consumption ($\mathcal{O}(1)$), as evidenced by the flat trajectory in Fig.~\ref{fig:memory}. This efficiency confirms that our recurrent inference formulation enables sequential iteration over WSI tiles without caching historical states, successfully resolving the trade-off between training efficiency and inference scalability. Notably, while PathRWKV exhibits linear time complexity ($\mathcal{O}(N)$) same with other methods, it maintains a competitive inference speed, effectively balancing its relatively large parameter size for complex feature modeling with computational efficiency.

To further assess the interpretability of our framework, Fig.~\ref{fig:cam} visualizes the Class Activation Maps (CAMs) for representative samples from the CAMELYON16 dataset. Note that, as standard attention maps are not directly obtainable from Prov-GigaPath and S4MIL, we visualize their saliency maps instead. Consistent with quantitative findings, SlideAve and SlideMax fail to generate meaningful activation patterns due to their simplistic aggregation logic. Compared with the ground truth, attention-based methods (e.g., ABMIL, CLAM) tend to assign uniform weights across the entire region of interest, demonstrating a limited capacity to distinguish fine-grained intratumoral heterogeneity. Furthermore, they exhibit relatively higher attention scores in normal regions. Among transformers, TransMIL misdirects attention to incorrect areas, while Prov-GigaPath focuses on normal regions in the second sample; this is likely attributable to overfitting on the small-scale dataset. In contrast, all SSMs successfully detect the accurate regions. Notably, PathRWKV not only accurately highlights global tumor regions but also delineates local feature intensity variances through its heatmap distribution. This visualization validates the effectiveness of the multi-scale modeling facilitated by the Time Mix and Channel Mix modules, demonstrating PathRWKV's ability to extract hierarchically significant pathological features.
\begin{figure*} [htbp]
  \centerline{\includegraphics [width=\textwidth]{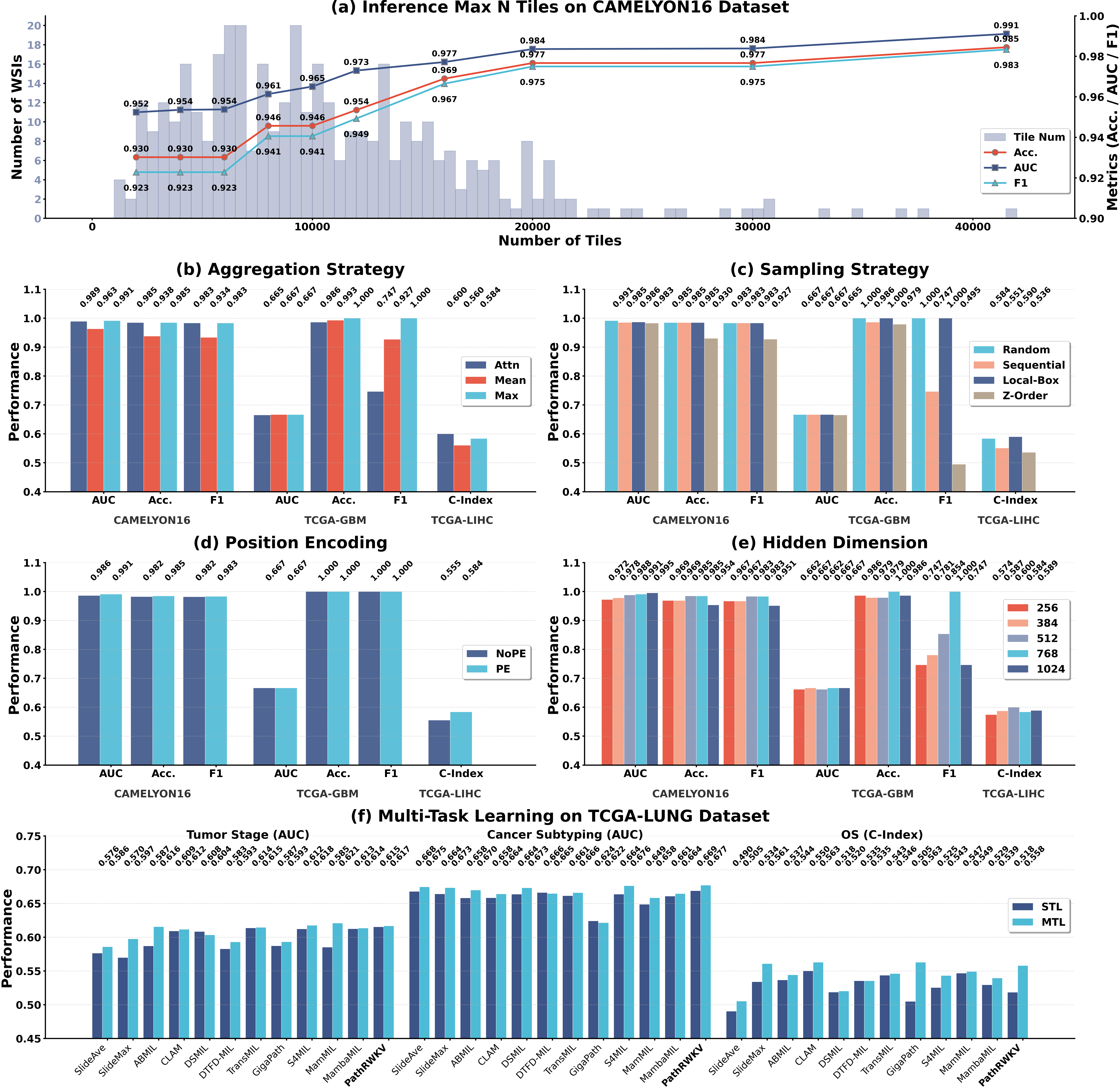}}
  \caption{Ablation studies and performance analysis. (a) Sensitivity analysis of inference performance with respect to the maximum number of input tiles on the CAMELYON16 dataset. (b)--(e) Impact of key components on model performance, including sampling strategies, aggregation methods, position encoding, and hidden dimension sizes across CAMELYON16, TCGA-GBM, and TCGA-LIHC datasets. The baseline strategies are consistently shown in cyan. (f) Comparison between Single-Task Learning (STL) and multi-task learning (MTL) on the TCGA-LUNG dataset for tumor staging, cancer subtyping, and overall survival prediction.}
  \label{fig:ablation}
\end{figure*}

\section{Discussion}
To rigorously evaluate the contribution of each component within PathRWKV and validate our design choices, we conducted a comprehensive series of ablation studies. The results are summarized in Fig.~\ref{fig:ablation}.

\subsection{Validation of Asymmetric Design and Aggregation}
A core innovation of PathRWKV is the asymmetric structure, designed to resolve the conflict between training throughput and inference memory efficiency. This design is predicated on two critical hypotheses: first, that a model trained on short sequences can effectively generalize to full-length WSIs during inference; and second, that max pooling serves as a sufficient and efficient aggregator for slide-level features. To validate these premises, we conducted ablation studies on each component.

\noindent \textbf{Inference Scalability.}
Our asymmetric protocol involves training on a fixed subsample (2,000 tiles) while inferring on the entire WSI (up to 40,000+ tiles). Fig.~\ref{fig:ablation}a analyzes the impact of inference sequence length on performance. Despite the potential distribution shift caused by the length discrepancy, PathRWKV exhibits a continuous performance improvement as the number of inference tiles increases. The steepest performance gains coincide with the peak of the WSI tile count distribution (approx. 8,000--12,000 tiles), indicating that the model effectively integrates information from the entire slide. This confirms that our recurrent backbone successfully captures long-range dependencies and generalizes well to sequence lengths far exceeding those seen during training.

\noindent \textbf{Aggregation Strategy.} 
The choice of aggregation function dictates both the representation quality and memory complexity. While Gated Attention (Attn) is the standard in MIL, it necessitates storing all tile features in memory to compute global softmax weights, leading to $\mathcal{O}(N)$ memory usage. In contrast, our proposed streaming max pooling strategy maintains $\mathcal{O}(1)$ complexity. 
As shown in Fig.~\ref{fig:ablation}b, max pooling demonstrates remarkable competitiveness. Compared to Attention, it achieves comparable performance on CAMELYON16 and substantially higher Accuracy and F1 scores on TCGA-GBM (Acc: 0.993 vs. 0.986; F1: 1.000 vs. 0.747). Crucially, max pooling aligns with the worst-pattern diagnostic principle in pathology, where the presence of a specific high-grade lesion often dictates the diagnosis, rendering the global average less relevant. However, we acknowledge the limitation that strictly selecting the maximum feature may discard information regarding tumor burden and the global microenvironment, which are valuable for survival analysis. This is reflected in the TCGA-LIHC task, where Attention slightly outperforms max pooling (0.600 vs. 0.584) by capturing global context. Nevertheless, given the massive efficiency gain ($\mathcal{O}(1)$ vs. $\mathcal{O}(N)$ memory), max pooling represents an optimal trade-off for efficient WSI modeling.

\subsection{Spatial-Temporal Robustness}
To address the overfitting risks on small-scale datasets and the loss of spatial structure due to sampling, we introduced random sampling paired with 2D PE.

\noindent \textbf{Sampling Strategy.}
Fig.~\ref{fig:ablation}c compares four sampling strategies. Sequential sampling feeds image features and coordinates according to their original extraction order, maintaining a deterministic sequence. Random sampling introduces a global random permutation, mitigating potential biases associated with the raster scanning order. Z-order sampling~\cite{peng2025one} utilizes a space-filling curve to retain 2D spatial locality within the flattened 1D sequence. Local-box sampling prioritizes dense local contexts by randomly selecting centroids and querying spatial neighbors. Intuitively, structure-preserving strategies like Sequential or Z-Order might seem superior for an RNN-based model like PathRWKV. However, the empirical results counter-intuitively favor random sampling, which achieves the highest metrics on CAMELYON16 and TCGA-GBM. We hypothesize that random sampling acts as a potent data augmentation technique, breaking the model's reliance on specific raster-scanning orders and preventing overfitting to incidental sequence patterns. While it disrupts local spatial continuity, it forces the model to learn more robust, permutation-invariant representations.

\noindent \textbf{2D Position Encoding.}
The efficacy of random sampling is intrinsically linked to the inclusion of 2D PE. As random sampling discards the implicit spatial order, PE is essential for explicitly injecting coordinate information back into the features. Fig.~\ref{fig:ablation}d corroborates this, showing that adding PE consistently maintains or enhances performance across tasks (e.g., boosting TCGA-LIHC C-Index from 0.555 to 0.584). This confirms that PathRWKV utilizes these encodings to reconstruct the spatial context of the tissue microenvironment, thereby mitigating the structural information loss caused by random sampling.

\subsection{Generalization and Optimization}
Finally, we analyzed the components designed to enhance model generalization under data-constrained conditions.

\noindent \textbf{Hidden Dimension.}
We investigated the impact of model capacity by varying the hidden dimension $D$ (Fig.~\ref{fig:ablation}e). Increasing $D$ does not linearly translate to better performance. While $D=1024$ yields the highest AUC on CAMELYON16, it degrades performance on the smaller TCGA-LIHC dataset, likely due to overfitting. Conversely, $D=768$ offers the optimal balance, achieving the highest performance on TCGA-GBM and competitive results elsewhere. This finding underscores the importance of matching model complexity to the scale of available pathological data, validating our choice of 768 as the default configuration.

\noindent \textbf{Multi-task Learning.}
The MTL module is designed to regularize feature learning by leveraging auxiliary tasks. Fig.~\ref{fig:ablation}f illustrates the performance of various backbones with and without MTL on the TCGA-LUNG dataset. The results demonstrate that our MTL module is a versatile plugin, improving the performance of most baselines (e.g., boosting ABMIL and CLAM). While there are marginal drops in specific cases for single-task specialists (e.g., a 0.4\% drop for DSMIL on Tumor Stage), the overall trend signifies that learning shared representations across related clinical tasks effectively reduces overfitting and improves the robustness of the slide-level features.

\noindent \textbf{Limitations.}
Despite the promising results, we acknowledge certain limitations in our current study. Specifically, there is a potential risk associated with the streaming max pooling strategy in other complex downstream clinical tasks, as it may discard subtle contextual cues. Furthermore, our current evaluation relies exclusively on Prov-GigaPath features, highlighting the need for future research to validate PathRWKV across a more diverse range of pathology foundation models.
\section{Conclusion}
In conclusion, this work proposed PathRWKV, a novel slide-level modeling framework that introduces an asymmetric training and inference paradigm, provides a principled solution to long-standing challenges in whole slide image analysis. 
To the best of our knowledge, PathRWKV is the first approach to explicitly decouple slide-level training and inference within a unified structure, enabling robust learning during parallelized training while preserving holistic slide reasoning at inference. Through the integration of asymmetric state space modeling, random tile sampling with multi-task learning regularization, 2D sinusoidal position encoding, and multi-scale feature mixing, PathRWKV effectively addresses weak supervision, data scarcity, disrupted spatial context, and heterogeneous multi-scale feature interactions. 
Extensive experiments on 29,073 whole slide images across 11 public datasets validating its effectiveness and reliability, demonstrating its strong potential to support real-world clinical workflows. 
We believe PathRWKV establishes a new perspective on slide-level modeling by showing that asymmetry between training and inference is not a limitation, but a powerful inductive principle for scalable and trustworthy computational pathology.
\section{Acknowledgments}
We gratefully acknowledge the contributions of several collaborators, particularly Haowen Hou from the National University of Singapore, for their invaluable support. We also extend our gratitude to Bo Peng from the RWKV Project under the Linux Foundation AI \& Data for their insightful suggestions and guidance. 
\bibliography{Sections/References}
\section{Theoretical Analysis of Asymmetric Structure}
\label{sec:theory}

In the main text, we introduce an asymmetric structure that utilizes parallel processing during training and recurrent state-passing during inference. Here, we provide the mathematical proof demonstrating that the chunk-based recurrent inference is mathematically equivalent to processing the entire WSI sequence in a single pass.

It is worth noting that despite the difference in execution modes, both approaches rely on the same fundamental linear operations without introducing any complex approximations. As revealed by the non-trivial derivations of the closed-form solution (Eq.~\eqref{eq:closed_form}) and the state-passing mechanism (Eq.~\eqref{eq:state}) below, the memory of the model is mathematically preserved through simple linear decays. This ensures that no global context information is lost due to the chunking strategy.

Recall the core state update rule of the PathRWKV block defined in Eq.~\eqref{eq:sequential} of the main text. For a sequence of tiles indexed by $t$, the hidden state matrix $S_t$ and the output $y_t$ are computed as:
\begin{align}
    S_t &= w_t \odot S_{t-1} + k_t v_t^\top \label{eq:update_rule} \\
    y_t &= r_t \odot (S_{t-1} + u \odot k_t v_t^\top) \label{eq:output_rule}
\end{align}
where $\odot$ denotes element-wise multiplication, and $w_t$ represents the data-dependent decay at step $t$. $S_0$ is initialized as a zero matrix.

\begin{proposition}[Associativity and Inference Equivalence]
\label{prop:equivalence}
Let $\mathcal{X} = \{x_1, x_2, \dots, x_N\}$ be the complete sequence of tiles from a WSI.
Let $\mathcal{A}$ denote the computation of the final state $S_N$ by processing all tiles continuously:
\begin{equation}
    S_N = \Phi(\mathcal{X}, S_0)
\end{equation}
Let $\mathcal{B}$ denote the computation where the sequence is split into two contiguous chunks $\mathcal{X}_1 = \{x_1, \dots, x_M\}$ and $\mathcal{X}_2 = \{x_{M+1}, \dots, x_N\}$ (where $1 < M < N$). The inference is performed sequentially by passing the intermediate state:
\begin{align}
    S_M &= \Phi(\mathcal{X}_1, S_0) \\
    S_N' &= \Phi(\mathcal{X}_2, S_M)
\end{align}
Then, the final states are identical: $S_N = S_N'$.
\end{proposition}

\begin{proof}
The recursive update rule in Eq.~\eqref{eq:update_rule} implies that the current state depends on the immediate past state, which in turn depends on its predecessor. By recursively unrolling this dependency back to the initial step $t=1$, we can observe a pattern: the contribution of an input $k_i v_i^\top$ at step $i$ to the current state $S_t$ is scaled by the cumulative product of all subsequent decay factors. Mathematically, this accumulation allows us to express $S_t$ in a non-trivial closed form:
\begin{equation}
\label{eq:closed_form}
    S_t = \sum_{i=1}^{t} \left( \prod_{j=i+1}^{t} w_j \right) \odot (k_i v_i^\top) + \left( \prod_{j=1}^{t} w_j \right) \odot S_0
\end{equation}
Assuming $S_0 = \mathbf{0}$, the term involving $S_0$ vanishes.

\noindent \textbf{Case 1: Global Continuous Inference ($\mathcal{A}$)}
Applying Eq.~\eqref{eq:closed_form} to the full sequence $t=N$:
\begin{equation}
    S_N = \sum_{i=1}^{N} \left( \prod_{j=i+1}^{N} w_j \right) \odot (k_i v_i^\top)
\end{equation}
We can split this summation into two parts at index $M$.
\begin{equation}
\label{eq:split_sum}
\begin{split}
    S_N &= \underbrace{\sum_{i=1}^{M} \left( \prod_{j=i+1}^{N} w_j \right) \odot k_i v_i^\top}_{\text{Part 1}} \\
        &+ \underbrace{\sum_{i=M+1}^{N} \left( \prod_{j=i+1}^{N} w_j \right) \odot k_i v_i^\top}_{\text{Part 2}}
\end{split}
\end{equation}
Notice that for Part 1, the decay product can be factored: $\prod_{j=i+1}^{N} w_j = (\prod_{j=M+1}^{N} w_j) \odot (\prod_{j=i+1}^{M} w_j)$.

\noindent \textbf{Case 2: Chunked Sequential Inference ($\mathcal{B}$)}
First, compute the state $S_M$ after the first chunk $\mathcal{X}_1$:
\begin{equation}
    S_M = \sum_{i=1}^{M} \left( \prod_{j=i+1}^{M} w_j \right) \odot k_i v_i^\top
\end{equation}
Next, use $S_M$ as the initial state for the second chunk $\mathcal{X}_2$. We apply the recursive definition starting from step $M+1$ to $N$. Here, we treat $S_M$ similarly to $S_0$ in Eq.~\eqref{eq:closed_form}, but with a crucial difference: the historical information carried by $S_M$ must continue to decay as it propagates through the new sequence from $M+1$ to $N$. By induction, the final state $S_N'$ comprises two components: the decayed history from the previous chunk and the accumulated information from the current chunk:
\begin{equation}
    \label{eq:state}
    S_N' = \left( \prod_{j=M+1}^{N} w_j \right) \odot S_M + \sum_{i=M+1}^{N} \left( \prod_{j=i+1}^{N} w_j \right) \odot k_i v_i^\top
\end{equation}
Substitute $S_M$ into the equation above:
\begin{equation}
\begin{split}
    S_N' &= \left( \prod_{j=M+1}^{N} w_j \right) \odot \left[ \sum_{i=1}^{M} \left( \prod_{j=i+1}^{M} w_j \right) \odot k_i v_i^\top \right] \\
         &+ \sum_{i=M+1}^{N} \left( \prod_{j=i+1}^{N} w_j \right) \odot k_i v_i^\top
\end{split}
\end{equation}
Distributing the decay term $\prod_{j=M+1}^{N} w_j$ into the summation bracket exactly reconstructs the Part 1 term from Eq.~\eqref{eq:split_sum}, and the second term is identical to Part 2.
\begin{equation}
    S_N' \equiv S_N
\end{equation}
Thus, chunked inference with state passing is mathematically exact to global inference.
\end{proof}

\section{Implementation Details and Hardware Acceleration}
\label{sec:implementation}

In this section, we provide a detailed description of the implementation of the PathRWKV backbone, specifically focusing on the custom CUDA kernels designed to enable the asymmetric training and inference structure described in Section~\ref{sec:methods}.

\subsection{Custom CUDA Kernels}
To efficiently implement the mathematical duality of the Linear Attention mechanism, we implemented two distinct sets of CUDA kernels, corresponding to the parallel (training) and recurrent (inference) modes.

\noindent \textbf{Parallel Kernel for Training.} During training, we utilize the \textbf{wkv6 parallel} kernel. This kernel is optimized for maximizing throughput when the entire sequence is available in memory. It implements the time-decayed aggregation described in Eq.~\ref{eq:parallel}. Crucially, it fuses the computation of receptance ($r$), key ($k$), value ($v$), and time-decay ($w$) processing into a single GPU kernel to minimize memory access overhead (HBM reads/writes). The backward pass kernel analytically computes gradients for all parameters, including the time-dependent decay rates, ensuring stable backpropagation through long sequences without the vanishing gradient problem typical of RNNs. The kernel leverages shared memory tiling and loop unrolling to accelerate the accumulation of attention scores along the sequence dimension $T$.

\noindent \textbf{State-based Kernel for Inference.} For inference on gigapixel WSIs, we utilize the \textbf{wkv6 state} kernel. This kernel explicitly manages the recurrent state to support the chunked processing strategy. Unlike standard attention kernels, this kernel accepts an additional input tensor $S_{in}$ (the hidden state from the previous chunk) and outputs $S_{out}$ (the updated state). This directly implements the update rule derived in Proposition~\ref{prop:equivalence}. To further accelerate inference, the state update loop is vectorized using \texttt{float4} data types, allowing the GPU to process 4 floating-point numbers simultaneously per thread. The kernel performs internal accumulation in \texttt{float32} to maintain numerical precision during the recursive updates, preventing error accumulation over long WSI sequences.

\subsection{Chunked Inference Implementation}
The code implementation orchestrates the interaction between the CUDA kernel and the WSI data to minimize memory footprint.
Let $N$ be the total number of tiles and $B_{chunk}$ be the chunk size. The inference process proceeds as follows:

\noindent \textbf{Initialization.} A state tensor $S$ of shape $(B, H, N_{head}, N_{head})$ is initialized to zeros, where $H$ is the number of heads and $N_{head}$ is the head dimension.

\noindent \textbf{Sequential Processing.} The WSI is split into $\lceil N / B_{chunk} \rceil$ chunks. For each chunk $k$:
    \begin{equation}
        Y_k, S_{k} = \mathrm{Block}(X_k, S_{k-1})
    \end{equation}
    Here, $\mathrm{Block}$ calls the \textbf{wkv6 state} CUDA kernel, and $S_k$ represents the hidden state after processing the $k$-th chunk. The state $S$ is passed strictly from CPU to GPU memory only once per chunk, minimizing PCI-e bandwidth usage.
    
\noindent \textbf{Streaming Aggregation.} Concurrently with feature extraction, the slide-level representation is updated using the streaming max pooling aggregation: $h_{global} = \max(h_{global}, \max(Y_k))$. This ensures the GPU memory usage remains constant $\mathcal{O}(1)$ regardless of slide size.
\end{document}